\documentclass[aps,prl,twocolumn,superscriptaddresst]{revtex4-2}
\usepackage[T1]{fontenc}
\usepackage{amsmath}
\usepackage{amssymb}
\usepackage{graphicx}
\usepackage{hyperref}
\usepackage{xcolor}
\renewcommand{\vec}[1]{\mathbf{#1}}

\begin{document}

\title{Mapping the Stability of
Spin Qubits in Superconducting  Pseudogap Systems}

\author{Chen-How Huang}
\affiliation{Department of Physics and Nanoscience Center, University of Jyväskylä, P.O. Box 35 (YFL), FI-40014 University of Jyväskylä, Finland}

\author{Miguel A. Cazalilla}
\affiliation{Donostia International Physics Center (DIPC), 20018 Donostia-San Sebastian, Spain}
\affiliation{Ikerbasque, Basque Foundation for Science, 48013 Bilbao, Spain}

\begin{abstract}
Superconducting spin qubits, realized as Yu-Shiba-Rusinov spin-doublet states in quantum-dot-superconductor systems, represent a cornerstone of current research in quantum technologies. We analyze these ground states of  quantum impurities in superconducting pseudogap systems, namely  systems with a  pseudogap tunneling density of states $\rho(\epsilon) \sim |\epsilon|^r$ for energies $|\epsilon|\gg \Delta$ ($\Delta$ being a $s$-wave pairing potential). For $r=1$, these hosts are realized as Dirac materials (graphene or 3D topological insulator surfaces) in proximity to conventional superconductors, or as $d+i s$ superconductors. Using  effective field theory and numerical renormalization group, we  map the  phase diagram against the pseudogap exponent $r > 0$ and particle-hole symmetry-breaking perturbations. At particle-hole symmetry, increasing  $r$ also increases the critical value, $J_c$, of the Kondo coupling that triggers 
the transition from spin doublet to  singlet. Unlike the gapless pseudogap Kondo systems, numerical and analytical evidence suggest that Andreev reflection stabilizes a singlet ground state at $J$ for all  $r > 0$. Breaking particle-hole symmetry ---by potential scattering or chemical potential--- eventually restores the transition at  lower $J_c$. Our results indicate that coupling to superconducting hosts with large pseudogap exponents enhances the stability of spin qubits at large Kondo coupling.
\end{abstract}
\date{\today}
\maketitle

\begin{figure}[hb]
\centering
\includegraphics[width=0.8\linewidth]{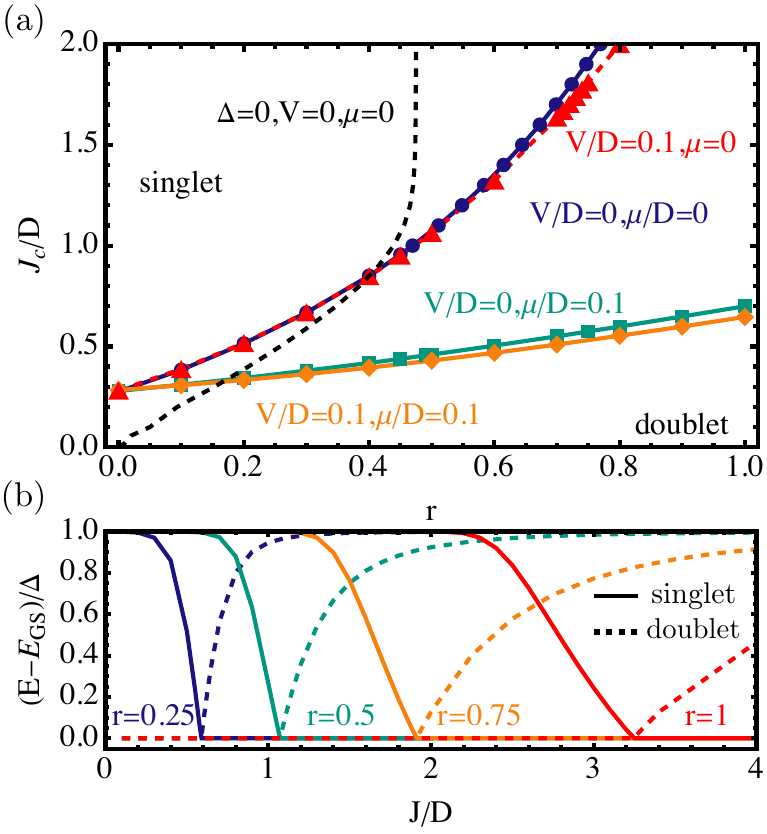}

    \caption{ Panel (a)  Ground state phase diagram for a magnetic impurity in
    a superconducting system with a pseudogap
    density of states $\sim |\epsilon|^r$ for quasi-particle energies $|\epsilon|\gg \Delta$ $\Delta$ being the $s$-wave pairing potential. $J_c/D$ is the ratio of the Kondo exchange coupling $J > 0$ and the system bandwidth  $D$. Different phase boundaries are displayed for different values of a local scattering potential $V$ chemical potential $\mu$, which break particle-hole  symmetry 
    Panel (b) Energy gap between the lowest-lying spin-singlet and doublet of opposite fermion parity. The gap closes at increasing values of $J_c/D$ as $r$ increases from zero.}
    \label{fig:fig1}
\end{figure}

\textit{\textcolor{blue}{Introduction:}} Superconducting spin qubits (SCSQ)~\cite{Nazarov_PhysRevB.81.144519,PitaVidal2023,Nazarov2_PhysRevLett.90.226806,pitavidal2025review} are rapidly becoming vital components for building practical, decoherence-free quantum memories within minimal Majorana chains~\cite{Leijnse2012,Sau2012,Deng2016,Dvir2023,Bordin2025}. These chains typically utilize arrays of quantum dots coupled to  superconductors (SCs) or semiconducting wires placed in close proximity to a SC~\cite{Leijnse2012,Sau2012}. Localizing a spin-$\tfrac{1}{2}$ at the dot is a fundamental ground-state property of the combined dot-SC system. In the absence of external magnetic fields~\cite{Leijnse2012,Sau2012} or spin-orbit coupling~\cite{Zazunov_PhysRevLett.103.147004,Sau2012,PitaVidal2023}, this configuration  is a degenerate spin doublet or local moment (LM).

The LM is ``erased'' by the occurrence of Kondo screening~\cite{Wilson_RevModPhys,Hewson_1993}. The latter is believed to inevitably happen as   the Kondo exchange of the dot $J$  is increased beyond a critical value, $J_c$, e.g. by making the dot-SC barrier(s)  more transparent 
deep in the Coulomb-blockade regime~\cite{Glazman_Pustilnik_2004}. Kondo screening leaves behind a spin-singlet ground state~\cite{Satori1992,SHIBA1993239,Bulla_RevModPhys.80.395}, which exhibits a much weaker response  to small  magnetic fields~\footnote{large magnetic fields suppress Kondo screening but are also detrimental to superconductivity.} or  spin-orbit coupling~\cite{Nozieres_1974,Hewson_1993}. The transition from  doublet to singlet  is indeed a level crossing transition,  which changes the ground-state fermion parity, $P$, as the doublet and singlet  have opposite $P$.  Close to the transition, the  singlet-doublet energy gap becomes small and the system is more susceptible to quasi-particle poisoning that causes   $P$ to  fluctuate~\cite{Aumentadoetal2023,Pan2022Engineering}.

In this work, we  address the following question: By modifying the host tunneling density of states (TDOS)  (or, equivalently, the energy dependence of the tunneling amplitude~\cite{Bulla_RevModPhys.80.395}),  how much can we increase  $J_c$? Specifically, we consider what we shall refer to as ``superconducting pseudogap systems'' (SCPS), namely 
hosts with a pseudogap TDOS  $\rho(\epsilon) \sim |\epsilon|^r$ ($r > 0$) at energies larger than the $s$-wave pairing potential $\Delta$. The answer to this question is relevant for the stability of a SCSQ in quantum dots coupled with SCPS. 
Examples of SCPS with $r = 1$ are  graphene~\cite{WangJ_PhysRevB.98.121411,Trivini2025} or topological insulators in proximity to conventional  SCs~\cite{FuKane2008,Stanescuetal2010,Wangetal2012,Xuetal2014,Loss_PhysRevB_2014,Trivini2025}, or (quasi-two dimensional) $d+i s$-wave SCs. We note that $d + i s$-wave pairing has recently been proposed as a  candidate to explain the  pairing symmetry of certain SC phases observed in infinite-layer nickelates~\cite{Gu2020,Ji_NatComm2023,Normand_PhysRevB.110.024514,Wang_PhysRevB.102.220501}.   

 Indeed, extending the approach of Yu~\cite{Yu}, Shiba~\cite{Shiba}, and Rusinov~\cite{Rusinov} (YSR) to SCPS, qualitatively suggests that the value of $J_c$ can be enhanced relative to the metallic case (see Refs.~\cite{Loss_PhysRevB_2014,unpub}). For instance, for $r=1$, $J_c/D\sim D/(\Delta \ln |D/\Delta|)\sim 10$
for $\Delta/D = 10^{-3}$ (versus $J_c/D \sim 1$ for $r =0$), 
 $D$ being the  bandwidth.
However, since YSR neglected quantum fluctuations of the impurity spin as well as
Kondo screening, such an estimate is not reliable. Here we show
that, although a pseudogap TDOS  with $r > 0$ removes  single-particle states  at low energies weakening  Kondo screening, the spin of the ground state (doublet or singlet)  at large $J$ is
determined by a rich interplay of different emergent  scales, which include Andreev reflection off the impurity and the polarization energy of the Kondo singlet.  Thus,  even in the simpler limit of particle-hole symmetry, the behavior of the system is different --albeit related-- to the well-studied pseudogap Kondo problem~\cite{Fradkin_PhysRevLett.64.1835,Chen_1995,Ingersent_PhysRevB.57.14254,Ingersent_PhysRevB.57.14254,Fritz_Vojta_2004}. The analytical results are supported by calculations using the numerical renormalization group, which allows us to  quantitatively map the ground state phase diagram  (cf. Fig.~\ref{fig:fig1}). 

As shown below, SCPS  allow for a substantial increase in
$J_c$ and the doublet-singlet gap at large Kondo coupling $J$. Enhancing the stability of SCSQs at large  $J$ has a number of potential advantages: A larger doublet-singlet gap improves  protection  against quasiparticle poisoning in a larger parameter regime. In addition, it allows to increase the strength and range of two-qubit interactions leading to e.g. more efficient two-qubit gates. Long-distance spin-spin interactions in hybrid superconducting systems  can be engineered using Rudermann-Kittel-Kasuya-Yoshida (RKKY)-type interactions~\cite{Choi_PhysRevB.62.13569,Leijnse_PhysRevLett.111.060501,Hassler_PhysRevB.92.235401} and 
other approaches~\cite{Nazarov2_PhysRevLett.90.226806,PitaVidal2023}. To lowest order, the coupling is proportional to the fourth power of the  tunneling amplitude into the dot,  and therefore $\propto J^2$.  In addition, as the chemical potential is tuned to neutrality, the RKKY interaction in superconducting graphene or topological insulators  falls off as \emph{pure} power-law and does not oscillate
with distance as it is the case of metals  (see e.g.~\cite{Loss_PhysRevB_2014}).

\textit{\textcolor{blue}{Model:}} We consider a quantum dot described as spin-$\tfrac{1}{2}$ quantum impurity (see below) in a SCPS. Superconductivity (or proximity to a $s$-wave superconductor) is described within the mean-field approach to the Bardeen-Cooper-Schrieffer (BCS) theory by a pairing potential,  $\Delta$. For $d+i s$ SCs, there is an additional $d$-wave pairing potential whose amplitude $\Delta_d \gg \Delta$. For TIs and graphene in general, 
$\Delta$ depends on the details of the SC/TI or SC/Graphene interface or device geometry. Here we will not be interested in the detailed form of $\Delta$ resulting from the proximity effect and therefore assume it to be constant. 
For devices in which the contact with the SC is lateral, this is a good approximation provided the distance of the impurity (quantum dot) to the SC electrode is  smaller  than the superconducting coherence length~\cite{Loss_PhysRevB_2014,Trivini2025}.  This may not be the case  in SC/TI(Graphene)/SC junctions due to multiple Andreev reflection~\cite{Bouchiat_PhysRevB.94.115405,WangJ_PhysRevB.98.121411}.

The dot spin is coupled to the host via a Kondo exchange $J>0$, which results from a Schrieffer-Wolff (SW)-like elimination~\cite{Hewson_1993,Glazman_Pustilnik_2004} of the dot levels. The average number electrons in the dot,  $\bar{N} = \langle \hat{N}\rangle$  is \emph{odd} and can tuned by gating. In the Coulomb blockade regime, the dot's charging energy, $E_C$, suppresses  fluctuations of $\hat{N}$ and thus the magnetic degrees of freedom of the dot can be described by the ground-state spin-$\tfrac{1}{2}$ operator, $\vec{S}$~\cite{Pustilnik_PhysRevLett.87.216601,Glazman_Pustilnik_2004}. The system  Hamiltonian (using Einstein's convention  for spin indices $\sigma =\uparrow,\downarrow$) reads:
\begin{align}
H &=  H_K + H_c, \label{eq:model}\\ 
H_{I}&=  \frac{\pi}{k_0 L}  \sum_{k,k^{\prime}}\, c^{\dagger}_{k\sigma} \left( J\vec{s}_{\sigma\sigma^{\prime}} \cdot \vec{S} + V \delta_{\sigma\sigma^{\prime}}\right)c_{k^{\prime}\sigma^{\prime}}, \notag \\
H_c &=  \sum_{k} \, \left[  \xi_{k}    c^{\dag}_{k\sigma}    c_{k\sigma} + 
  \left( \Delta c^{\dagger}_{k\uparrow} c^{\dagger}_{-k,\downarrow}  + \mathrm{H.c.} \right)\right].\notag
\end{align}
Here the SC host is effectively represented as a  one-dimensional  (chiral) system of length $L$  which describes the channel the with largest Kondo coupling to the dot spin~\cite{Hewson_1993,Pustilnik_PhysRevLett.87.216601,Glazman_Pustilnik_2004}. The (quasi-particle) Fermi operators
$c_{k\sigma}, c^{\dag}_{k\sigma}$ ($\sigma = \uparrow,\downarrow$)  obey
$\{ c_{k\sigma},c^{\dagger}_{k^{\prime}\sigma^{\prime}} \} = \delta_{k,k^{\prime}} \delta_{\sigma,\sigma^{\prime}}$, and anti-commute otherwise; $\vec{s} = \tfrac{1}{2}(\sigma^x, \sigma^y, \sigma^z)$, $\sigma^{\alpha= x,y,z}$ being the  Pauli matrices. 
For $\Delta = 0$, quasi-particles
have power-law dispersion $\xi_k 
= \epsilon_k - \mu$, where $\epsilon_k = v k | k/k_0|^{-r/(1+r)}$  ($r\ge 0$, with $r = 0$ for constant DOS and $r = 1$ for Dirac systems);
$v$ has units of velocity and $k_0$ is a momentum cutoff related to the system bandwidth $D = v k_0$; The  chemical potential is $\mu$; for $\mu = 0$, the 
dispersion is particle-hole symmetric and yields a pseudogap DOS $\rho(\epsilon) =  \rho_0  \left| \frac{\epsilon}{D}\right|^r$ with $r > 0$;   $\rho_0 = (r+1)/2D$,  $D$ being the bandwidth. For the $d+is$-wave SCs considered here,  $D\sim \Delta_d\gg \Delta$~\cite{unpub}. Furthermore, since the $d$-wave nodal points are pinned to the Fermi surface,  $\mu = 0$. 

 In addition to the Kondo coupling, we have assumed  a local scattering potential of strength $V$. Both $V$ and $\mu$  break  particle-hole symmetry (PHS); $V$ stems from elastic co-tunneling processes through the dot levels as well as  the particle-hole asymmetry of band structure. In quantum dots $V$ accounts for additional effects that are not captured by the standard SW treatment of the Anderson model~\cite{Pustilnik_PhysRevLett.87.216601,Glazman_Pustilnik_2004}. For  $ \Delta\neq 0$, an SW-type elimination generates additional terms which can be neglected near PHS~\cite{Salomaa_PhysRevB.37.9312}, i.e. near the middle of the Coulomb blockade valley where $V\approx 0$ and $J = 4 \sum_{n} t^*_n  \Lambda_{mn}t_m/E_C$, $t_{n}$ being the tunneling amplitude of the strongest coupled channel into the dot's $n$'th  level, and $\Lambda_{mn}$ a matrix that characterizes the ground-state projection of the dot spin onto  $\vec{S}$~\cite{Pustilnik_PhysRevLett.87.216601,Glazman_Pustilnik_2004}. 

\textit{\textcolor{blue}{Stability of SCSQ at weak Kondo coupling:}} At small $J$, the stability of the spin-doublet (SCSQ) as ground state of \eqref{eq:model}  can be studied using  perturbative renormalization group (RG)~\cite{Fradkin_PhysRevLett.64.1835,Hewson_1993}. By eliminating high-energy  states, $j \propto \rho_0 J$ is renormalized according to
\begin{equation}
\frac{d j}{d \ln D} = r j - j^2 + \cdots \label{eq:rg1}
\end{equation}
Renormalization proceeds as long as $D$ can be decreased, i.e. for $D \gtrsim \Delta$ since below the  gap there are no available states to be integrated out. Solving
 \eqref{eq:rg1} for $r > 0$ up to  $D \approx \Delta$ shows that the bare $J(D) = J \ll D$ decreases to $J(\Delta) \approx J (\Delta/D)^r$. Crudely, if $J(\Delta) \ll \Delta$, BCS pairing favors a singlet  ground state and the dot's spin effectively decouples, which yields a spin-doublet (LM) ground state. Thus,  weak coupling, the stability of SCSQs in SCPS increases with  $r$.  

\textit{\textcolor{blue}{Stability of SCSQ at large Kondo coupling:}} At large $J$,
the fate of the LM  is  tied to the possibility of  Kondo singlet (KS) formation. 
The low-energy properties of the KS can be described by  effective  theories that  differ from the ``microscopic'' model, Eq.~\eqref{eq:model}. In Ref.~\cite{unpub}, we provide a  pedestrian
derivation the theory that applies to SCPS close to PHS (i.e. for $\mu = V = 0$ in Eq.~\ref{eq:model}).  
We first consider the system exactly at PHS, with deviations from it being discussed further below. The effective action is local in space and non-local in Matsubara time,  and can be written as follows:
 \begin{align}
 S &= S_0 + S_1, \label{eq:sK}\\ 
 S_0 &= \frac{1}{\beta} \sum_{\omega_n} \bar{\Psi}_c G^{-1}_c(\omega_n) \Psi_c,\label{eq:sKSC}\\
 G^{-1}_c(\omega_n) &= A_0 D_r(\omega_n) \left(i\omega_n-\tau^1 \Delta \right)  \notag \\
 &+ i A_1 \omega_n + \tau^1  B_1  \Delta  \label{eq:ginvc},\\
 D_r(\omega_n) &= 
\left\{ 
\begin{array}{cc}
\frac{\pi}{\cos(\pi r/2)}  \left(  \frac{\omega^2_n + \Delta^2}{\Lambda^2_K}\right)^{(r-1)/2}  & |r| < 1,\\
\ln\left|\frac{\Lambda_K^2}{\omega^2_n + \Delta^2} \right|   & r = 1.
\end{array}
\right. \\
S_1 &= U_0\int d\tau \left(\bar{c}_{\uparrow} c_{\uparrow} - \tfrac{1}{2} \right) \left( \bar{c}_{\downarrow} c_{\downarrow} - \tfrac{1}{2} \right) + \cdots \label{eq:hubbard}
 \end{align}
 where  $\Psi_c = \left( c_{\uparrow}, \bar{c}_{\downarrow} \right)^T$ and $\bar{\Psi}_c = \left( \bar{c}_{\uparrow}, c_{\downarrow} \right)$ are Nambu spinors,  $c_{\sigma},\bar{c}_{\sigma}$  are  Grassmann fields that describe the system local degrees of freedom after integrating out the host and the KS. In a Wilson-chain description~\cite{Wilson_RevModPhys,Hewson_1993,Bulla_RevModPhys.80.395}, the $c$-fermions describe the low-energy dynamics of the 2nd site of the chain after eliminating all the other sites, including the 1st site  that is effectively frozen out in KS with the impurity. The Matsubara frequencies 
fulfill  $|\omega_n| < \Lambda_K \lesssim D$, where $\Lambda_K$ is a high energy cutoff (e.g. for $r = 0$, $\Lambda_K \sim T_K$, i.e. the Kondo temperature~\cite{Hewson_1993}). The $2\times 2$ matrices $\tau^{n=1,2,3}$ are the Pauli matrices in Nambu space.  In the strong coupling regime where $J\gg D$, the  term $\propto A_1$ perturbatively arises from singlet-doublet transitions of the KS. The quadratic action \eqref{eq:sKSC} describes  the mean-field saddle-point at $N\to +\infty$ of an SU($N$) generalization of Eq.~\eqref{eq:model}~\cite{Fradkin_PhysRevLett.64.1835,Read_1983,Huang_Scipost_2024},
 after integration of the host degrees of freedom. The additional Hubbard-like interaction $\propto U_0$ is related to  singlet-triplet fluctuations, i.e. the polarization of the KS and in the large-$N$ approach is  captured by  $1/N$ corrections~\cite{Read_1983}. For constant TDOS ($r=0$) and $\Delta = 0$, Eq.~\eqref{eq:sK} is effectively a path integral formulation of Nozi\'eres' local Fermi liquid theory~\cite{Nozieres_1974,Hewson_1993}, whose generalization to superconducting systems with $r=0$ was discussed in Ref.~\cite{Egger_PhysRevLett.121.207701}.

Having introduced the low-energy effective theory in Eq.~\eqref{eq:sK},   we shall briefly summarize  
a few important results regarding the RG flow near the KS  fixed point (named symmetric strong coupling or SSC in~\cite{Ingersent_PhysRevB.57.14254}) for $\Delta = 0$.  
For the  analysis of Eq.~\eqref{eq:sK}, we require the term $\propto A_0$ to be invariant under RG transformations. This amounts to assuming that the SSC fixed-point  is  a resonant-level model with the level pinned at $\epsilon = 0$~\cite{Fritz_Vojta_2004}). To leading order the RG flow  is controlled by the scaling dimensions of $A_1$ and $U_0$ (the term $\propto B_1$ vanishes at $\Delta = 0$), which are $r-1$, and $2r-1$, respectively~\cite{Fritz_Vojta_2004}. Hence,   $A_1$ and $U_0$ are both irrelevant for $r < \tfrac{1}{2}$ meaning that SSC is stable and large $J$.  For  $\tfrac{1}{2} < r < 1$, the  term $\propto U_0$ is a relevant perturbation that destabilizes the SSC~\cite{Chen_1995,Ingersent1_PhysRevB.54.11936,Fritz_Vojta_2004}. Roughly speaking, the flow of $U_0$ to strong coupling favors the single occupancy of the $c$-fermion site, resulting in a LM ground state. The SSC  merges with an intermediate-coupling critical point (SCR)~\cite{Fritz_Vojta_2004}, thereby causing all RG flows for $r > \tfrac{1}{2}$ to terminate at the spin-doublet LM fixed point.  
Indeed, NRG shows~\cite{Chen_1995,Ingersent_PhysRevB.57.14254}  that the LM  is  the only stable ground state for $r > \tfrac{1}{2}$ at PHS  (see  dashed line in Fig.~\ref{fig:fig1}).

 In view of these results, it is tempting to conclude that, for a small $\Delta\ll \Lambda_K$ at PHS, a spin singlet is never a stable ground state for $r > \frac{1}{2}$ and large $J$. However, this expectation is not correct because for $\Delta \neq 0$, the action~\eqref{eq:sK}
contains an additional perturbation $\propto B_1 \Delta \tau^1$. This local pairing term is called Andreev reflection (AR) (see~\cite{Egger_PhysRevLett.121.207701,Andersen_PhysRevLett.107.256802} for $r=0$).  Its effect on the SSC fixed point can be assessed by means of RG. Technically this requires approximating the first term in Eq.~\eqref{eq:ginvc} by 
its asymptote $\sim i A_0 \mathrm{sgn}( \omega_n) \left| \omega_n\right|^r$ for $|\omega_n|\gg \Delta$ and integrating out high-energy states  only up to $\Lambda_K\approx \Delta$. To leading order, the flow is controlled by the scaling dimension of $B_1$, which is $r$, meaning that AR is the most relevant perturbation to SSC  for $r > 0$. Indeed, as recognized in~\cite{Fritz_Vojta_2004} for a different physical realization,  AR has the same  dimension as a PHS-breaking perturbation~\cite{Fritz_PhysRevB_2005}, i.e. a scattering potential~\cite{unpub}, which in the Nambu notation is $\tau^3 V_1$. This is not surprising because, for $|\omega_n|\gg \Delta$,  we can rotate $\tau^1 B_1$  into a  $ \tau^3 V_1 $ by means of a Bogoliubov transformation.  Similar to potential scattering~\cite{Ingersent1_PhysRevB.54.11936,
Ingersent_PhysRevB.57.14254,Fritz_PhysRevB_2005,Fritz_Vojta_2004}, AR drives the RG flow away from SSC and favors  a spin-singlet ground state  where the $c$-fermions  are paired  (for our gauge choice~\cite{unpub}, either 
$|BCS\rangle = \left( |0\rangle - |2\rangle\right)/\sqrt{2}$ or $|\overline{BCS}\rangle = \left( |0\rangle + |2\rangle\right)/\sqrt{2}$,  depending on the sign of $B_1$, where $|0\rangle$ is the empty $c$-site and $|2\rangle = c^{\dag}_{\uparrow}c^{\dag}_{\downarrow}|0\rangle$).

\begin{figure}[t]
 \centering
 \includegraphics[width=0.8\linewidth]{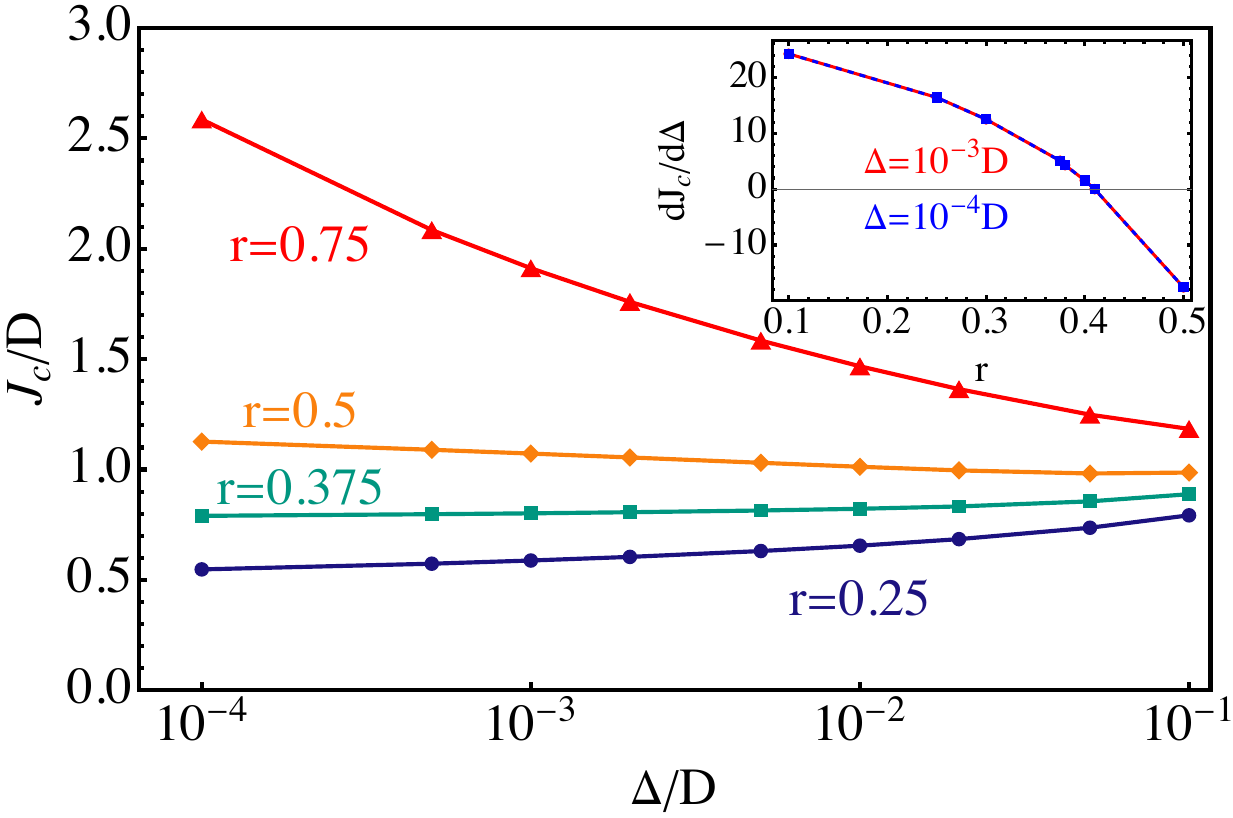}
    \caption{Dependence of the critical Kondo coupling $J_c$ on the strength of the ($s$-wave) pairing potential $\Delta/D$ ($D$ is the system bandwidth) at particle-hole symmetry (i.e. $V = \mu = 0$). Note that for $r = 0.75$, larger $\Delta$ favors a singlet ground state. Inset: Slope of the critical $J$, i.e. $dJ_c/d\Delta$ vs. pseudogap exponent $r$. Notice the sing change for $r\simeq 0.4$.} 
    \label{fig:fig2}
\end{figure}

The singlet ground state described above is a paired, particle-hole rotated version of the asymmetric strong-coupling  (ASC) fixed point discussed in~\cite{Ingersent_PhysRevB.57.14254,Fritz_PhysRevB_2005,Fritz_Vojta_2004} for $\Delta = 0$. Below we shall  refer to it as ``paired strong coupling'' (PSC). In Refs.~\cite{Ingersent_PhysRevB.57.14254,Fritz_PhysRevB_2005,Fritz_Vojta_2004}, for general $J$ and $V$,  the details of the RG flow between the LM and ASC fixed points depend on $r$ and the strength of  PHS breaking  (i.e. $V$)~\cite{Ingersent_PhysRevB.57.14254,Fritz_Vojta_2004}. However,  for a gapped system,   quantum critical points like SCR  or  the particle-hole  asymmetric  ACR~\cite{Ingersent_PhysRevB.57.14254,Fritz_Vojta_2004} are not, strictly speaking, accessible. Thus, 
for the most part, we shall refrain from  analyzing  the effects of quantum criticality on the system properties (see discussion below regarding the inset in Fig.~\ref{fig:fig2}, nonetheless). Instead, we  focus on the RG fixed points describing phases: At PHS and for $0 < r < \tfrac{1}{2}$, the stability of SSC in the $U_0$  direction  ``funnels''  RG flows into the PSC, leading to a crossover from the SSC  to a PSC singlet at large $J$.  On the other hand, for $r > \tfrac{1}{2}$, the competition between the relevant couplings  $B_1, (V_1)$ and $U_0$  can be crudely described by a single-site model for the $c$-fermion site~\cite{unpub}.  At the phase boundary, there is a level crossing  where the $c$-fermion site fluctuates between three states: $|\sigma \rangle = c^{\dag}_{\sigma}|0\rangle$ ($\sigma = \uparrow,\downarrow$) and e.g. $|BCS\rangle = (|0\rangle - |2\rangle)/\sqrt{2}$~\cite{unpub},  reminiscent of the valence-fluctuation (VFI) critical point that controls the LM/ASC transition for $r\ge 1$, $V\neq 0$, and $\Delta =0$~\cite{Ingersent_PhysRevB.57.14254,Fritz_Vojta_2004}. Accounting for a finite PHS-breaking scattering potential $V$ (i.e. a additional term $\propto V_1\tau^3$ in Eq.~\eqref{eq:sKSC}) does not modify  the above discussion substantially, provided the PSC state is defined by an appropriate a Bogoliubov rotation depending on the renormalized values of $B_1$ and $V_1$~\cite{unpub}.

Finally, we briefly consider the effect of the chemical potential, $\mu$. The latter  also breaks PHS and leads to a crossover from the $r > 0$ to $r = 0$ case, since for $\epsilon  = \mu$ and $\Delta =0$ the TDOS is
constant. Therefore,  an effective model like~\eqref{eq:model} with $r=0$, a reduced effective bandwidth $D^*\sim |\mu|$, and renormalized  $J^*$ and $V^*$ can be considered instead. Thus, the crossover-induced renormalization of $J, V\to J^*,V^*$ can enhance  $J_c$ at small $|\mu|$.

\textit{\textcolor{blue}{Numerical Renormalization Group (NRG) results:}}
 In order to  quantitatively obtain the boundary between the doublet and singlet ground states, we use NRG~\cite{Wilson_RevModPhys,Hewson_1993,Satori1992,Bulla_RevModPhys.80.395}. In Fig.~\ref{fig:fig1}(a), we show the phase diagram as a function of the ratio of the critical Kondo coupling to the bandwidth, $J_c/D$, and the pseudogap exponent, $r$. Besides results at PHS (i.e. $V = \mu = 0$),  we  have  also show the computed phase boundary for several combinations of $V\neq 0$ and $\mu\neq 0$. 
 
   The dashed line in Fig.~\ref{fig:fig1} is the phase boundary for the gapless system (i.e.
   $\Delta = 0$) at PHS, which was  computed in Ref.~\cite{Ingersent_PhysRevB.57.14254} using NRG. The latter is a continuous quantum phase transition. The boundary slope becomes  vertical as $r\to \tfrac{1}{2}^{-}$, implying that the spin singlet is never the ground state for  $r > \tfrac{1}{2}$ at PHS.  Turning  on a finite $\Delta = 10^{-3} D$, we observe a doublet-singlet transition  for $r > \tfrac{1}{2}$ at large $J$, consistent with discussion above based on effective field theory.  For  $r=1$ , which is relevant for  $d+is$ SCs and graphene or TIs in proximity to SCs, the transition is not visible in Fig.~\ref{fig:fig1}(a), but the  doublet-singlet level crossing happening at $J_c/D\simeq 3.2$ is shown in Fig.~\ref{fig:fig1}(b) along with  the doublet-singlet gap vs. $J/D$ for  several other  $r$ values. This plot shows a strong  trend for $J_c/D$ to increase with increasing $r$. In systems without PHS (i.e. $V\neq 0$ and/or $\mu\neq 0$)  a small $V$ has a very weak effect on the phase boundary. On the other hand, a finite $\mu$ has a dramatic effect, making the boundary  very weakly dependent on $r$, which  yields, at most, a factor of two enhancement in $J_c/D$ for $r=1$ relative to $r =0$.     

   Fig.~\ref{fig:fig2} illustrates the effect  on $J_c$ of $\Delta$, which controls the strength of AR (cf. Eq.~\ref{eq:sKSC}) for  $r \in \{ 0.25, 0.375,0.50, 0.75\}$:  $J_c$ is weakly dependent on $\Delta$ for $r \lesssim 0.5$. However, for $r=0.75$ NRG finds  a much stronger dependence, with the singlet ground state clearly being favored at large $\Delta$. The inset shows that $dJ_c/d\Delta$  changes sign near $r\simeq 0.4$. This is close to $r^*\simeq 0.375$ (for $r > r^{*}$, the ACR rather than the SCR  controls the transition from LM to ASC in the case of a PHS-breaking perturbation~\cite{Ingersent1_PhysRevB.54.11936,Fritz_Vojta_2004}).

\textcolor{blue}{Conclusions:} We have mapped the stability of spin-doublet or  spin qubit-type ground states in quantum dot-superconducting pseudogap systems using a effective field theory describing the  Kondo singlet ground state at large Kondo coupling $J$ and numerical renormalization group. For pseudogap exponent $r > 1/2$, Andreev reflection (AR) and the Kondo-singlet polarization compete, with AR favoring a paired-type of singlet ground state at large $J$ and the Kondo-singlet polarization favoring the spin-doublet.   The largest critical values of $J_c$ shown in Fig.~\ref{fig:fig1}(b) may not be accessible for quantum dots 
coupled to hosts such as graphene or topological insulator surfaces in proximity to a superconductor,  for which $r=1$ and $(J_c/D)_{\mathrm{max}} \simeq 3.2$. Thus the ground state will be a doublet in the full operational range. This will provide a route to engineer superconducting spin qubits that can be operated at large $J$ and are more resilient to quasi-particle poisoning. 

\textcolor{blue}{Acknowledgments:} MAC thanks S. Bergeret and G. Zarand for useful discussions. CHH and MAC have been been supported by the Spanish
MCIN/AEI/10.13039/501100011033 through Grant No. PID2023-148225NB-C32 (SUNRISE).
\bibliography{references}

\end{document}